\documentclass[final,5p,times,twocolumn]{elsarticle}
\usepackage[margin=0pt,font=small,labelfont=bf,labelsep=colon]{caption}
\usepackage{url}
\usepackage{hyperref}
\usepackage{amsmath}
\usepackage{cases}
\usepackage{natbib}
\usepackage{amssymb}
\usepackage{gensymb} %for degree sign
\usepackage{graphicx}
\usepackage{color}

\biboptions{sort&compress}
\journal{Carbon}

\begin{document}

\begin{frontmatter}

\title{Kapitza thermal resistance across individual grain boundaries in graphene}

\author[a,b]{Khatereh Azizi}
\author[b]{Petri Hirvonen}
\ead{petri.hirvonen@aalto.fi}
\author[b]{Zheyong Fan\corref{author}}
\author[b]{Ari Harju}
\author[c]{Ken R. Elder}
\author[b,d]{Tapio Ala-Nissila}
\author[a,e]{S. Mehdi Vaez Allaei}
\ead{smvaez@ut.ac.ir}

\cortext[author] {Corresponding author. \textit{E-mail:} brucenju@gmail.com (Zheyong Fan)}

\address[a]{Department of Physics, University of Tehran, Tehran 14395-547, Iran}
\address[b]{COMP Centre of Excellence, Department of Applied Physics,
Aalto University School of Science, P.O. Box 11000, FIN-00076 Aalto, Espoo, Finland}
\address[c]{Department of Physics, Oakland University, Rochester, Michigan 48309, USA}
\address[d]{Departments of Mathematical Sciences and Physics, Loughborough University, Loughborough,
 Leicestershire LE11 3TU, UK}
\address[e]{School of Physics, Institute for Research in Fundamental
Sciences (IPM), Tehran 19395-5531, Iran}

\begin{abstract}
We study heat transport across individual grain boundaries in suspended monolayer graphene using 
extensive classical molecular dynamics (MD) simulations. We construct bicrystalline graphene samples containing 
grain boundaries with symmetric tilt angles using the two-dimensional phase field crystal method and then relax the
samples with MD. The 
corresponding Kapitza resistances are then computed
using nonequilibrium MD simulations. We find that the Kapitza resistance depends strongly on the tilt angle 
and shows a clear correlation with the average density of defects in a given grain boundary, but is not strongly correlated with
the grain boundary line tension. We also show that quantum effects are significant in quantitative determination of 
the Kapitza resistance by applying the mode-by-mode quantum correction to the classical MD data.
The corrected data are in good agreement with quantum mechanical Landauer-B\"utticker calculations.
\end{abstract}


\end{frontmatter}

\section{Introduction}
\label{Introduction}

Graphene \cite{novoselov2004}, the famous two-dimensional  allotrope of carbon, has been demonstrated to have 
extraordinary electronic \cite{castro2009}, mechanical \cite{lee2008}, and thermal \cite{balandin2008} properties in its pristine form. 
However, large-scale graphene films, which are needed for industrial applications are typically grown by chemical vapor 
deposition \cite{li2009} and are polycrystalline in nature \cite{huang2011}, consisting of domains of pristine graphene 
with varying orientations separated by grain boundaries (GB) \cite{yazyev2010prb,liu2011carbon,liu2011nl}. They play a 
significant or even dominant role in influencing many properties of graphene \cite{yazyev2014,cummings2014}.

One of the most striking properties of pristine graphene is its extremely high heat conductivity, which has been shown to be in excess of 
$5000$ W/mK \cite{balandin2008,ma2017}. Grain boundaries in graphene act as line defects or one-dimensional interfaces which leads to a strong
reduction of the heat conductivity in multigrain samples \cite{hahn2016carbon,fan2017nl}. The influence of GBs can be quantified
by the Kapitza or thermal boundary resistance $R$. The Kapitza resistance of graphene grain boundaries has been previously computed using molecular dynamics (MD) \cite{bagri2011,cao2012} and Landauer-B\"utticker \cite{lu2012,serov2013} 
methods, and has also been measured experimentally \cite{yasaei2015}. However, these works have only 
considered a few separate tilt angles, and a systematic investigation on the dependence of the Kapitza resistance on the tilt angle 
between any two pristine grains is still lacking. The relevant questions here concern both the magnitude $R$
for different tilt angles and possible correlations between the structure or line tension of the GBs and the corresponding value of $R$.

Modelling realistic graphene GBs has remained a challenge due to the multiple length and time scales involved. 
Recently, an efficient multiscale approach \cite{hirvonen2016} for modelling polycrystalline graphene samples was developed based on 
phase field crystal (PFC) models \cite{elder2002,elder2004}. The PFC models are a family of continuum methods for modelling the 
atomic level structure and energetics of crystals, and their evolution at diffusive time scales (as compared to vibrational time scales in MD).
The PFC models retain full information about the atomic structure and elasticity of the solid \cite{elder2004}. 
It has been shown \cite{hirvonen2016} that using the PFC approach in two-dimensional space one can obtain large, realistic and 
locally relaxed microstructures that can be mapped to atomic coordinates for further relaxation in three-dimensional space with the usual atomistic simulation methods.

In this work, we employ the multiscale PFC strategy of Ref. \cite{hirvonen2016}  to generate large samples of tilted, bicrystalline graphene
with a well-defined GB between the two grains. These samples are then further relaxed with MD at $T = 300$ K. A heat current is generated
across the bicrystals using nonequilibrium MD (NEMD) simulations, and the Kapitza resistance is computed from the temperature drop across the GB.
We map the values of $R(\theta)$ for a range of different tilt angles $\theta$ and demonstrate how $R$ correlates with the structure of the GBs.
Finally, we demonstrate that quantum corrections need to be included in $R$ to obtain quantitative agreement with
experiments and lattice dynamical calculations.

\section{Models and Methods}
\label{section:Theory}

\subsection{PFC models}

PFC approaches typically employ a classical density field $\psi\left(\boldsymbol{r}\right)$ to describe the systems. The ground state of $\psi$ is 
governed by a free energy functional $F\left[\psi\left(\boldsymbol{r}\right)\right]$ that is minimized either by a 
periodic or a constant $\psi$, corresponding to crystalline and liquid states, respectively. We use the standard PFC model
\begin{equation}
F = \int d\boldsymbol{r} \left(\frac{1}{2}\psi\left[\epsilon+\left(q^2+\nabla^2\right)^2\right]\psi + \frac{1}{3}\tau\psi^3 + \frac{1}{4}\psi^4\right),
\end{equation}
where the model parameters $\epsilon$ and $\tau$ are phenomenological parameters related to temperature and average density, respectively. 
The component $\left(q^2+\nabla^2\right)^2$ penalizes for deviations from the length scale set by the wave number $q$, giving rise to a spatially 
oscillating $\psi$ and to elastic behaviour \cite{elder2002,elder2004}. The crystal structure in the ground state is dictated by the 
formulation of $F$ and the average density of $\psi$, and for certain parameter values the ground state of $\psi$ displays a 
honeycomb lattice of density maxima as appropriate for graphene \cite{hirvonen2016}.

\begin{figure*}
\centering
\includegraphics[width=13cm]{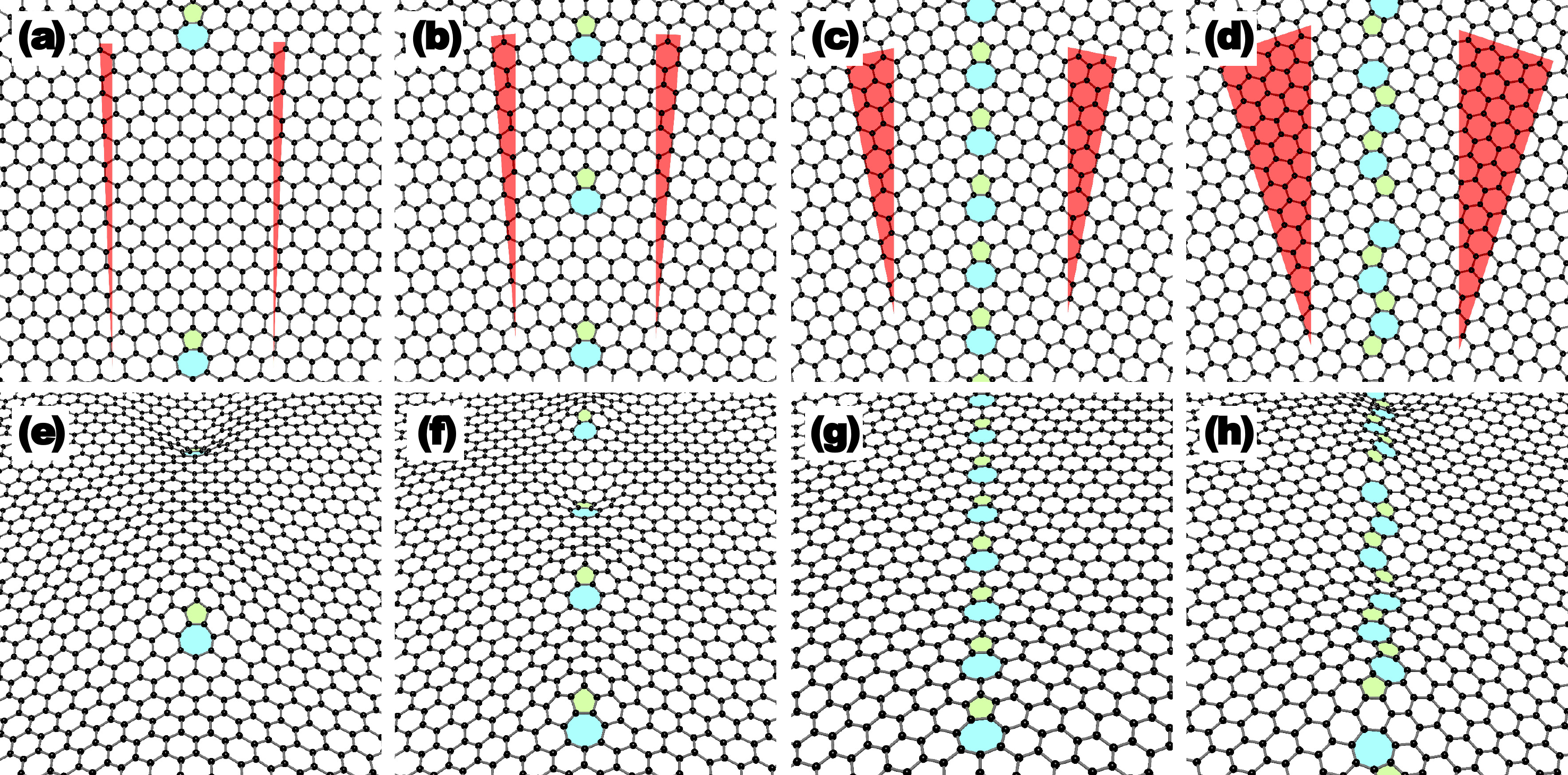}
\caption{A schematic illustration of the definition of the tilt angle in bicrystalline graphene with symmetrically tilted grain boundaries.
(a)-(d): Grain boundary with tilt angles (indicated by the red wedges) of $2\theta=$ $4.42\degree$, 
$9.43\degree$, $21.79\degree$, and $36.52\degree$ before MD relaxation. (e)-(h): the corresponding systems after MD relaxation. For clarity, only the small part close to the grain boundary is shown for each system; the sample size in our NEMD simulations is much larger, which is $L_x=400$ nm (transport direction) and $L_y=25$ nm (transverse direction).  (A colour version of this figure can be viewed online.)}
\label{fig_theta}
\end{figure*}

The PFC calculations are initialized with symmetrically tilted $2$-crystals in a periodic, two-dimensional 
computational unit cell. 
The initial guess for the crystalline grains is obtained by using the one-mode approximation \cite{elder2004}
\begin{equation}
\psi\left(x,y\right) = \cos{\left(qx\right)}\cos{\left(\frac{qy}{\sqrt{3}}\right)}-\frac{1}{2}\cos{\left(\frac{2qy}{\sqrt{3}}\right)},
\end{equation}
and by rotating alternatingly by $\pm\theta$. The tilt angle between two adjacent grains is $\theta-\left(-\theta\right) = 2\theta$, which ranges from 
$2\theta = 0\degree$ to $2\theta = 60\degree$ (see Fig. \ref{fig_theta} for examples). We consider a subset of the tilt angles investigated in Ref. \cite{hirvonen2016}, with the exact values being listed in Table \ref{table1}. The rotated grains and the unit cell size are 
matched together as follows: if just one of the rotated grains filled the whole unit cell, it would be perfectly continuous at the periodic edges. 
Along both interfaces, narrow strips a few atomic spacings wide are set to the average density -- corresponding to a disordered state -- to give the 
grain boundaries some additional freedom to find their lowest-energy configuration. We assume non-conserved dynamics to 
relax the systems in analogy to chemical vapour deposition \cite{kim2011}
-- the number of atoms in the monolayer can vary as if due to exchange with a vapor phase. In addition, the unit cell dimensions are 
allowed to vary to minimize strain. Further details of the PFC calculations can be found in Ref. \cite{hirvonen2016}. 
The relaxed density field is mapped to a discrete set of atomic coordinates suited for the initialization of MD simulations \cite{hirvonen2016}.

\subsection{NEMD simulations}

We use the NEMD method as implemented in the GPUMD (graphics processing units molecular dynamics) code \cite{fan2013,fan2015,fan2017cpc} to 
calculate the Kapitza resistance, using the Tersoff \cite{tersoff1989} potential with optimized parameters \cite{lindsay2010} for graphene.  The initial structures obtained by the PFC method are rescaled by an appropriate factor to have zero in-plane stress at 300 K in the MD simulations with the optimized Tersoff potential \cite{lindsay2010}.

\begin{figure}[ht]
\centering
\includegraphics[width=7cm]{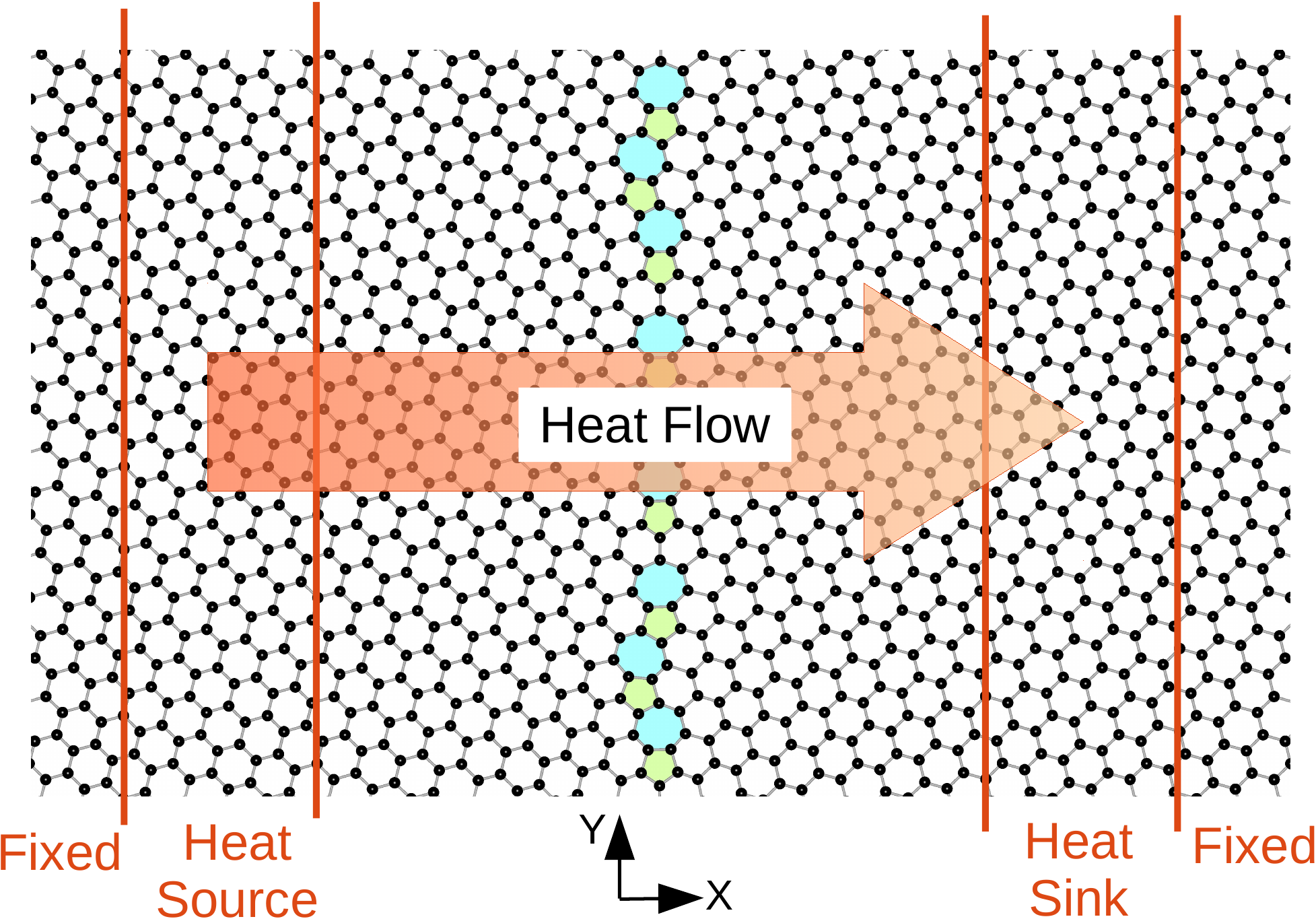}
\caption{A schematic illustration of the NEMD setup used in computing the Kapitza resistance.  (A colour version of this figure can be viewed online.)}
\label{fig:setup}
\end{figure}

In the NEMD simulations, periodic boundary conditions are applied in the transverse direction, whereas fixed boundary conditions are applied in the transport direction. We first equilibrate the system at 1 K for 1 ns, then increase the temperature from 1 K to 300 K during 1 ns, 
and then equilibrate the system at 300 K for 1 ns. After these steps, we apply a Nos\'{e}-Hoover chain of thermostats 
\cite{nose1984,hoover1985,martyna1992} to the heat source and sink, choosing as 
two blocks of atoms around the two ends of the system, as schematically shown in Fig. \ref{fig:setup}. The temperatures of the heat source and sink are maintained at 310 K and 290 K, respectively. We have checked that steady state 
can be well established within 5 ns. In view of this, we calculate the temperature profile $T(x)$ of the system and the energy 
exchange rate $Q$ between the system and the thermostats using data sampled in another 5 ns. The velocity-Verlet integration scheme \cite{swope1982} with a time step of 1 fs is used for all the calculations. 
Three independent calculations are performed for each system and the error estimates reported in Table \ref{table1} correspond to the standard error of the independent results.

\begin{figure}[ht]
\centering
\includegraphics[width=7cm]{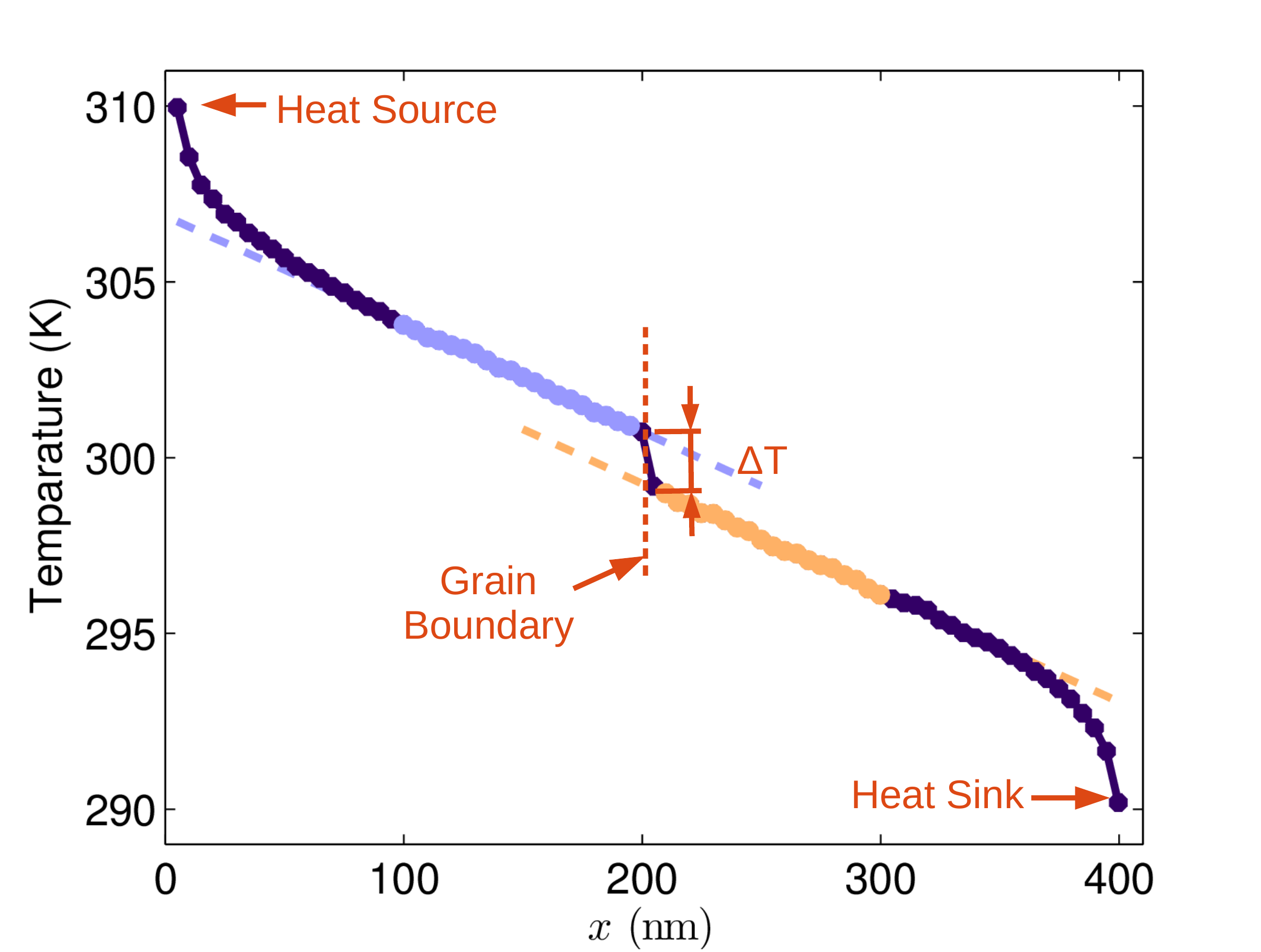}
\caption{ A typical steady-state temperature profile in 
bicrystalline graphene with a tilt angle of $2\theta=9.43\degree$. On each side of the grain boundary, excluding the nonlinear region around the heat 
source or the sink, one can fit the temperature by a linear function and then extract the temperature jump as the difference 
between the two linear functions at the grain boundary \cite{landry2009,bagri2011,rajabpour2011jap,cao2012,gordiz2014jap,bohrer2017}.  
(A colour version of this figure can be viewed online.)
}
\label{fig:T}
\end{figure}

In steady state, apart from the nonlinear regions around the heat source and the sink intrinsic to the method, 
a linear temperature profile can be established on each side of the GB, but with an inherent discontinuity (temperature jump) 
at the GB. An example of this for the system with $2\theta = 9.43\degree$ is shown in Fig. \ref{fig:T}. The Kapitza resistance $R$ is 
defined as the ratio of the temperature jump $\Delta T$ and the heat flux $J$ across the grain boundary:
\begin{equation}
R = \frac{\Delta T}{J},
\label{eq_G_NEMD}
\end{equation}
where $J$ can be calculated from the energy exchange rate $Q$ (between the system and thermostat) 
and the cross-sectional area $S$ (graphene thickness is chosen as $0.335$ nm in our calculations), {\it i.e.} $J=Q/S$.


\section{Results and Discussion}
\label{sec_results}

\begin{table*}[ht]
\centering
\caption{\label{table1} The GB tilt angle $2\theta$, the corresponding temperature jump $\Delta T$, 
heat flux $J$, Kapitza resistance $R$, Kapitza length $L_K$, grain boundary line tension $\gamma$, and defect density $\rho$ for the 13 bicrystalline graphene samples considered here.}
\begin{tabular}{rrrrrrr}
\hline
\hline
	$2\theta$ ($\degree$) & 
	$\Delta T$ (K) & 
	$J$ (GW/m$^{2}$) &
	$R$ (m$^2$K/GW) &
	$L_K$ (nm) &
	$\gamma$ (eV/nm) &
	$\rho$ (1/nm) \\
	\hline
  1.10 & $ 0.09 \pm  0.07$ & $51.3 \pm  0.4$ & $ 0.0018 \pm  0.0013$  & $ 10 \pm   7$   &  0.55   &  0.08 \\ 
  4.41 & $ 0.73 \pm  0.14$ & $50.4 \pm  0.6$ & $ 0.0144 \pm  0.0026$  & $ 75 \pm  13$   &  2.21   &  0.31 \\ 
  9.43 & $ 1.36 \pm  0.15$ & $45.7 \pm  0.3$ & $ 0.0298 \pm  0.0033$  & $155 \pm  17$   &  3.84   &  0.67 \\ 
 13.17 & $ 1.62 \pm  0.08$ & $48.1 \pm  0.5$ & $ 0.0337 \pm  0.0020$  & $175 \pm  10$   &  4.71   &  0.93 \\ 
 18.73 & $ 1.99 \pm  0.06$ & $43.9 \pm  0.2$ & $ 0.0453 \pm  0.0015$  & $236 \pm   8$   &  5.02   &  1.32 \\ 
 21.79 & $ 1.97 \pm  0.01$ & $47.8 \pm  0.2$ & $ 0.0412 \pm  0.0004$  & $214 \pm   2$   &  4.69   &  1.54 \\ 
 27.80 & $ 2.39 \pm  0.04$ & $43.6 \pm  0.3$ & $ 0.0548 \pm  0.0005$  & $285 \pm   3$   &  4.71   &  1.95 \\ 
 32.20 & $ 2.49 \pm  0.10$ & $43.4 \pm  0.7$ & $ 0.0574 \pm  0.0033$  & $298 \pm  17$   &  3.77   &  2.25 \\ 
 36.52 & $ 2.48 \pm  0.14$ & $43.9 \pm  0.5$ & $ 0.0565 \pm  0.0026$  & $294 \pm  14$   &  4.93   &  1.91 \\ 
 42.10 & $ 2.30 \pm  0.02$ & $43.1 \pm  0.2$ & $ 0.0534 \pm  0.0006$  & $278 \pm   3$   &  5.50   &  1.46 \\ 
 46.83 & $ 1.93 \pm  0.06$ & $46.4 \pm  0.3$ & $ 0.0416 \pm  0.0014$  & $216 \pm   7$   &  5.16   &  1.06 \\ 
 53.60 & $ 1.01 \pm  0.06$ & $41.2 \pm  0.3$ & $ 0.0245 \pm  0.0014$  & $127 \pm   7$   &  3.36   &  0.52 \\ 
 59.04 & $ 0.17 \pm  0.07$ & $47.3 \pm  0.1$ & $ 0.0036 \pm  0.0014$  & $ 19 \pm   7$   &  0.61   &  0.08 \\ 
 \hline
 \hline
\end{tabular}
\end{table*}
\begin{figure}[ht]
\centering
\includegraphics[width=7cm]{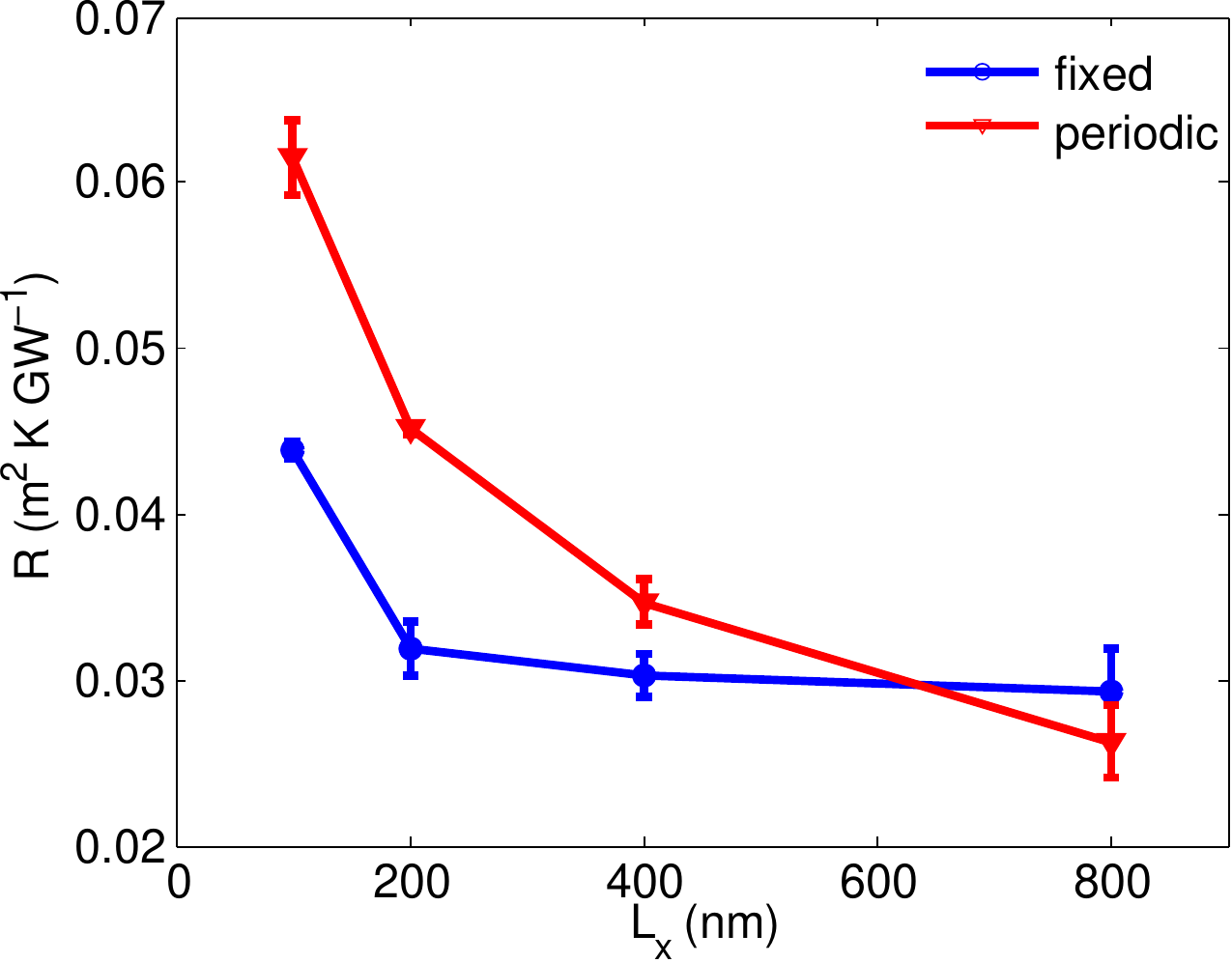}
\caption{Kapitza resistance as a function of the sample length $L_x$ obtained by using fixed boundary conditions (circles) and
periodic boundaries (triangles) in the case of $2\theta=9.43\degree$.
(A colour version of this figure can be viewed online.)
}
\label{fig_length_effect}
\end{figure}
\begin{figure}[ht]
\centering
\includegraphics[width=7cm]{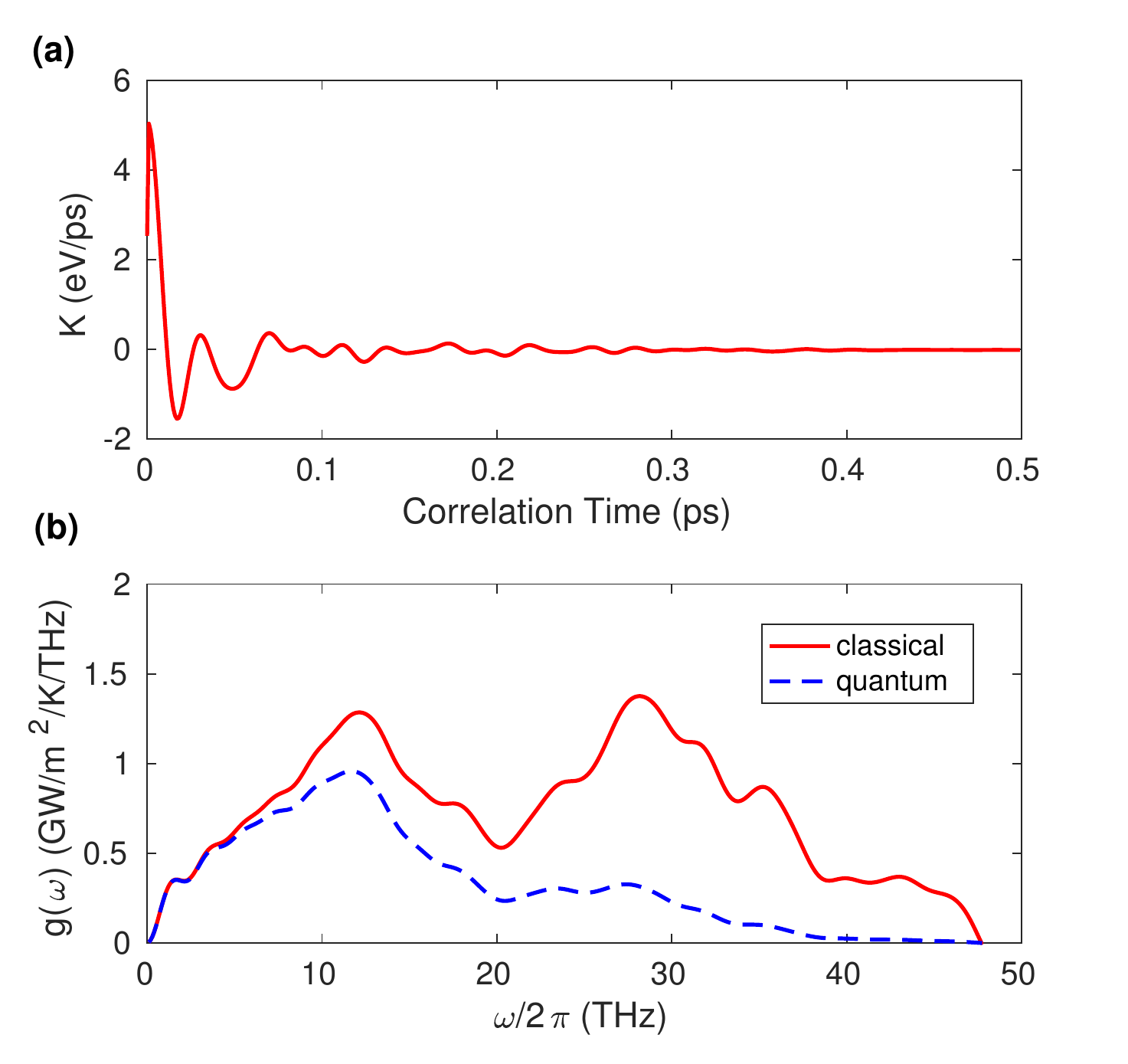}
\caption{(a) The nonequilibrium heat current correlation function $K(t)$ as a function of correlation time. (b) The spectral conductance before (solid line) 
and after (dashed line) mode-to-mode quantum corrections as a function of the phonon frequency. 
The system considered here corresponds to the case of $2\theta=9.43\degree$, but similar results are obtained for all other cases. 
(A colour version of this figure can be viewed online.)
}
\label{fig_quantum_effect}
\end{figure}

It is well known \cite{landry2009,bagri2011,cao2012} that the calculated Kapitza resistance depends on the sample length in 
NEMD simulations. Figure \ref{fig_length_effect} shows the calculated Kapitza resistance $R$ in the $2\theta=9.43\degree$ case as a function of the 
sample length $L_x$. Using fixed boundary conditions as described above, $R$ saturates at around $L_x=400$ nm. On the other hand, 
using periodic boundaries as described in Ref. \cite{bagri2011}, $R$ converges more slowly. To this end, we have here 
used fixed boundary conditions and a sample length of $400$ nm for all the systems. The 
calculated temperature jump $\Delta T$, heat flux $J$, and Kapitza resistance $R$ in the 13 bicrystalline systems are listed in Table \ref{table1}.

The Kapitza resistance calculated from the heat flux does not contain any information on the contributions from individual phonon modes. 
Methods of spectral decomposition of both the heat current (flux) \cite{saaskilahti2014,saaskilahti2015,saaskilahti2016,zhou2015a,zhou2015b,fan2017prb} 
and the temperature \cite{feng2017} within the NEMD framework have been developed recently. Here, we use the spectral decomposition formalism 
as described in Ref. \cite{fan2017prb} to calculate the spectral conductance $g(\omega)$ of the $2\theta=9.43\degree$ system. In this method, 
one first calculates the following nonequilibrium heat current correlation function ($t$ is the correlation time):
\begin{equation}
K(t) = \sum_{i\in A}\sum_{j\in B} 
\left\langle
\frac{\partial U_i(0)}{\partial \vec{r}_{ij}} \cdot \vec{v}_j(t) -
\frac{\partial U_j(0)}{\partial \vec{r}_{ji}} \cdot \vec{v}_i(t)
\right\rangle,
\end{equation}
where $U_i$ and $\vec{v}_i$ are respectively the potential energy and velocity of particle $i$, $\vec{r}_{ij}=\vec{r}_{j}-\vec{r}_{i}$ ($\vec{r}_i$ is the position of particle $i$), and $K(t=0)$ measures the heat current flowing form a block $A$ to an adjacent block $B$ arranged along the transport direction. 
Then, one performs a Fourier transform to get the spectral conductance:
\begin{equation}
g(\omega) = \frac{2}{S\Delta T} \int_{-\infty}^{+\infty} dt e^{i\omega t} K(t).
\end{equation}
The spectral conductance is normalized as
\begin{equation}
G= \int_{0}^{\infty} \frac{d\omega}{2\pi}  g(\omega),
\end{equation}
where $G$ is the total Kapitza conductance (also called thermal boundary conductance), which is the inverse of the Kapitza resistance $G=1/R$.

Figure \ref{fig_quantum_effect}(a) shows the calculated correlation function $K(t)$, which resembles the velocity autocorrelation function whose 
Fourier transform is the phonon density of states \cite{dickey1969}. Indeed, thermal conductance in the quasi-ballistic regime is intimately related 
to the phonon density of states. The corresponding spectral conductance $g(\omega)$ is shown as the solid line in Fig. \ref{fig_quantum_effect}(b). 
The total thermal boundary conductance is $G \approx 33$ GW/m$^{2}$/K, corresponding to a Kapitza resistance of $R \approx 0.03$ m$^2$K/GW.

In view of the high Debye temperature (around $2000$ K) for pristine graphene \cite{pop2012}, we expect that it is necessary to 
correct the classical results to properly account for possible quantum effects. While using classical statistics can lead to \cite{turney2009} 
an underestimate of the scattering time for the low-frequency phonons as well as an overestimate of the heat capacity of the high-frequency 
phonons for thermal transport in the diffusive regime, only the second effect matters here in the quasi-ballistic regime. 
Therefore, one can correct the results by 
multiplying the classical spectral conductance by the ratio of the quantum heat capacity to the classical one: 
$x^2e^x/(e^x-1)^2$, where $x=\hbar\omega/k_BT$, with $\hbar$, $k_B$, $T$ being the Planck constant, Boltzmann constant, and 
system temperature, respectively. This factor is unity in the low-frequency (high-temperature) limit and 
zero in the high-frequency (low-temperature) limit. Applying this mode-to-mode quantum correction to the classical 
spectral conductance gives the quantum spectral conductance represented by the dashed line in Fig. \ref{fig_quantum_effect}(b). 
The integral of the quantum corrected spectral conductance is reduced by a factor of about $2.3$ as compared to the classical one.  

\begin{figure}[ht]
\centering
    \includegraphics[width=7cm]{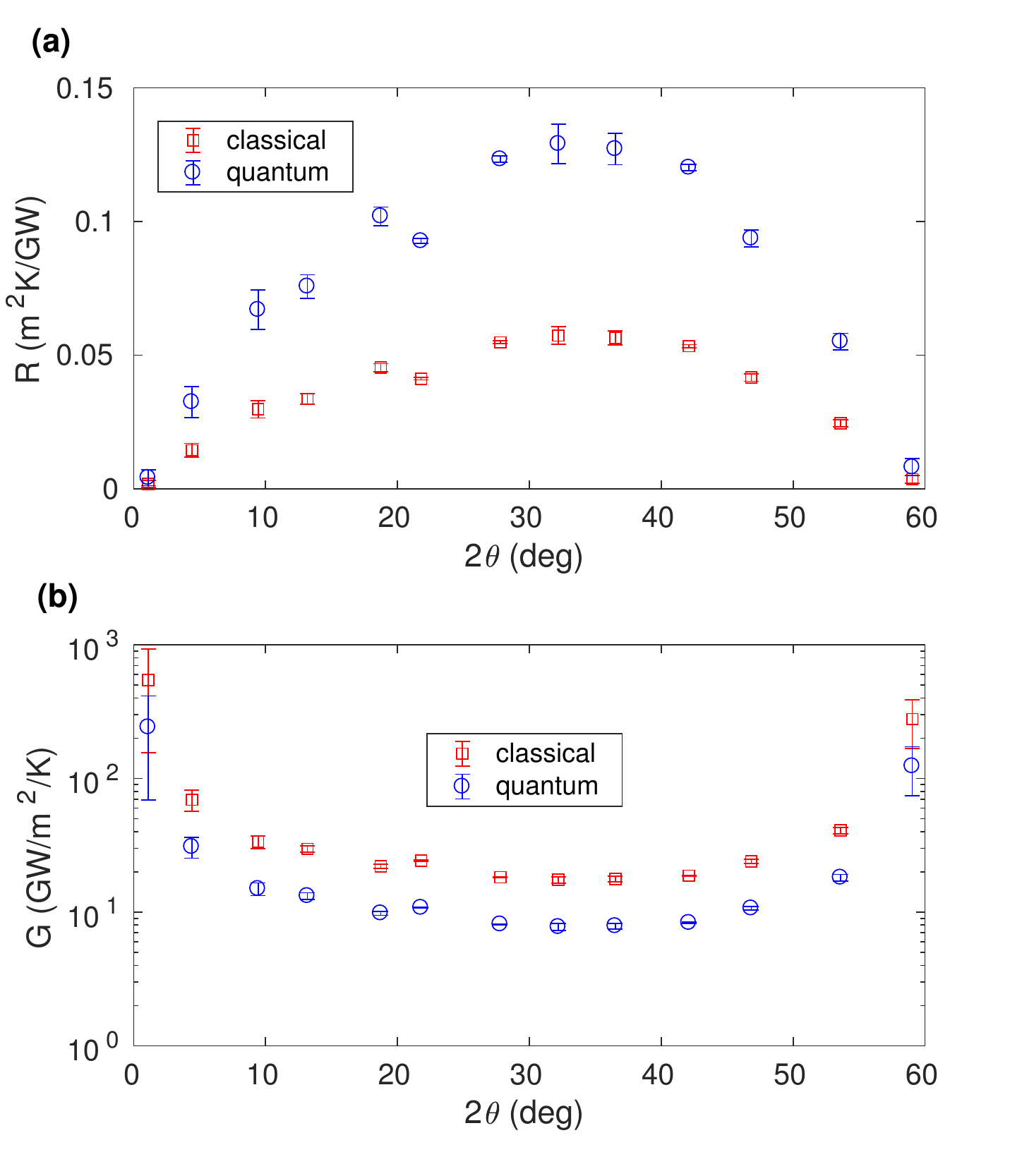}
\caption{(a) Kapitza resistance $R$ of the grain boundary before (labeled as ``classical'') and after (labeled as ``quantum'') the mode-to-mode quantum 
correction as a function of the tilt angle $2\theta$. (b) The corresponding Kapiza conductance $G$ as a function of the tilt angle.
 (A colour version of this figure can be viewed online.)
 }
\label{fig:RG}
\end{figure}

Figure \ref{fig:RG}(a) shows the calculated Kapitza resistances for all the $13$ systems as a function of the tilt angle, 
both before (red squares) and after (blue circles) applying the quantum correction. It clearly shows that the Kapitza 
resistance depends strongly on the tilt angle, varying by more than one order of magnitude. The Kapitza resistance increases monotonically from both sides to the middle angle of $2\theta \sim 30\degree$, except for one ``anomalous'' system with $2\theta=21.79\degree$. This system has smaller $R$ than that with $2\theta=18.73\degree$. One intuitive explanation is that this system is relatively flat compared to other systems, as can been seen from Figs. \ref{fig_theta}(e)-(h). Similar ``anomalous'' heat transport has been reported in Ref. \cite{liu2014carbon} for the same grain boundary tilt angle.

The largest Kapitza resistances 
occurring around the intermediate angles, being about $0.12$ m$^2$/K/GW after quantum corrections, 
are more than an order of magnitude smaller than those in grain boundaries in silicon nanowires \cite{bohrer2017}. A more reasonable comparison between different materials is in terms of the Kapitza length $L_K$ \cite{nan1997jap}, defined as the system length of the corresponding pristine material at which the bulk thermal resistance due to phonon-phonon scattering equals the Kapitza resistance. Mathematically, we have 
\begin{equation}
L_K=\kappa R,
\end{equation}
where $\kappa$ is the thermal conductivity of the bulk material. We calculate $L_K$ by assuming a value of $\kappa=5200$ W/mK for pristine graphene according to the very recent experiments \cite{ma2017} and list the values in Table \ref{table1}. The largest Kapitza lengths (corresponding to the largest Kapitza resistances) before quantum corrections are about 300 nm, which would be about 700 nm after quantum corrections. These values are actually larger than those for silicon nanowires. Therefore, the effect of grain boundaries on heat transport in graphene is not small even though the Kapitza resistances are relatively small.

To facilitate comparison with previous works, we also show the Kapitza conductances in Fig. \ref{fig:RG}(b). The Kapitza conductances in our 
systems range from about 17 GW/m$^{2}$/K to more than 500 GW/m$^{2}$/K before applying the quantum corrections. 
Bagri \textit{et al.} \cite{bagri2011} reported Kapitza conductance values (obtained by NEMD simulations with periodic boundary conditions in the transport direction) ranging from 15 GW/m$^{2}$/K to 45 GW/m$^{2}$/K. 
The lower limit of 15 GW/m$^{2}$/K does not conflict with our data, as this value is reported in a system of a grain size of 25 nm, 
where the data cannot have converged yet. On the other hand, Cao and Qu (obtained by NEMD simulations with fixed boundary conditions in the transport direction) \cite{cao2012} reported saturated Kapitza 
conductance values in the range of $19-47$ GW/m$^{2}$/K, which fall well within the values that we obtained. 
Last, we note that quantum mechanical Landauer-B\"utticker calculations by Serov \textit{et al.} \cite{serov2013} predicted the 
Kapitza conductance to be about 8 GW/m$^{2}$/K for graphene grain boundaries 
comparable to those in our samples with intermediate tilt angles ($2\theta \sim 30\degree$). This is much smaller than the classical 
Kapitza conductances (about 20 GW/m$^{2}$/K), but agree well with our quantum corrected values. This comparison justifies the
mode-to-mode quantum correction we applied to the classical data and resolves the discrepancy between the results from classical 
NEMD simulations and quantum mechanical Landauer-B\"utticker calculations.

\begin{figure}[ht]
\centering
    \includegraphics[width=7cm]{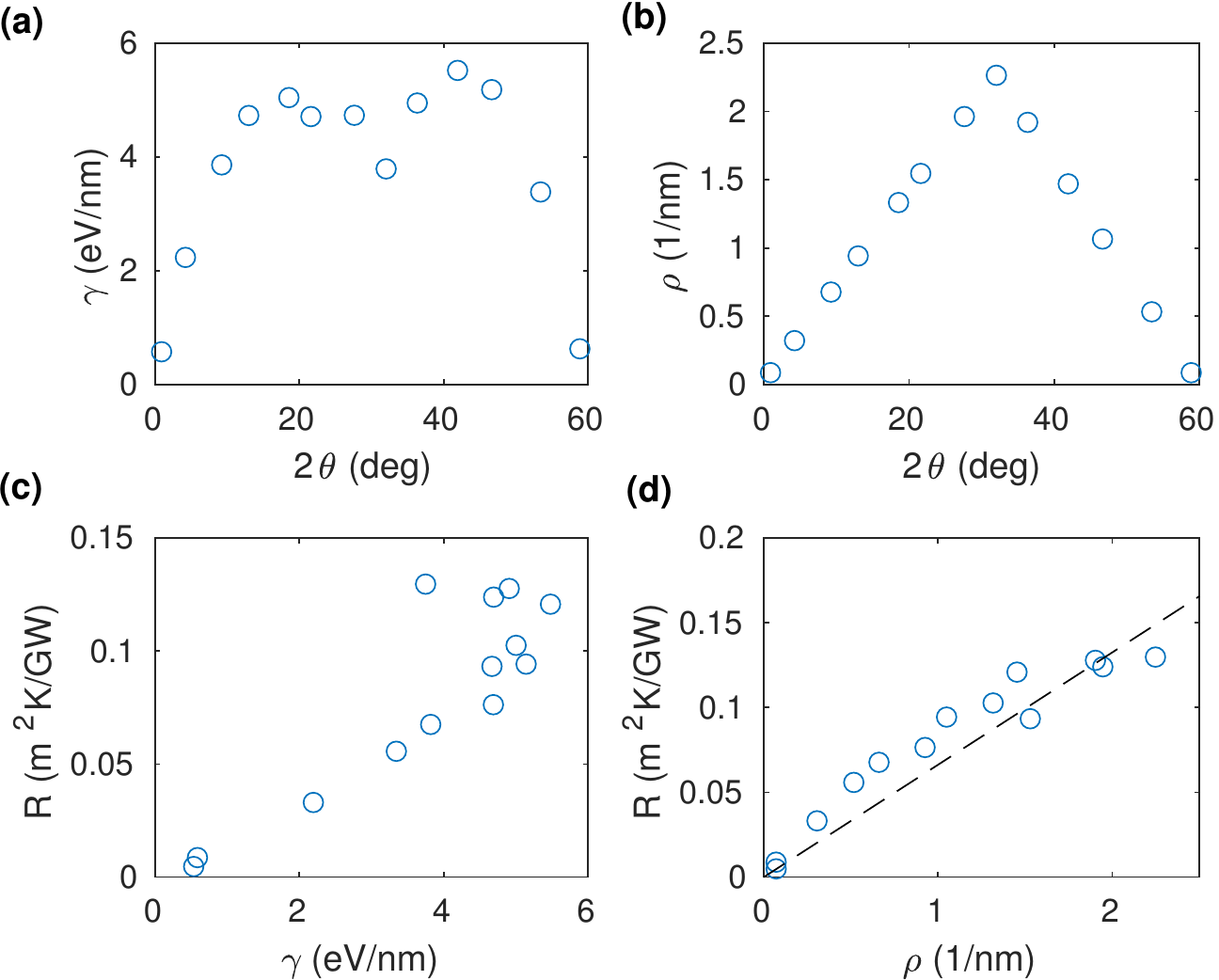}
\caption{(a) The grain boundary line tension $\gamma$ and (b) the defect density $\rho$ versus the tilt angle $2\theta$. (c) and (d) The quantum corrected Kapitza resistance $R$ versus $\gamma$ and $\rho$, respectively. The dashed line in (d) is a guide to the eye. (A colour version of this figure can be viewed online.)}
\label{fig:correlation}
\end{figure}

The last remaining issue concerns the possible correlation of the values of $R(\theta)$ with the energetics and structure of the GBs.
The grain boundary line tension and the defect density are closely related to the tilt angle. The line tension $\gamma$ is defined as
\begin{equation}
\gamma=\lim_{L_y \rightarrow \infty} \frac{\Delta E}{L_y},
\end{equation}
in the thermodynamic limit, where $\Delta E$ is the formation energy for a GB of length $L_y$. The defect density is defined as
\begin{equation}
\rho=\frac{N_{\rm p-h}}{L_y},
\end{equation}
where $N_{\rm p-h}$ is the number of pentagon-heptagon pairs in the grain boundary. The calculated $\gamma$ and $\rho$ 
values for all the tilt angles are listed in Table \ref{table1} and plotted in Figs. \ref{fig:correlation}(a)-(b). In Figs. \ref{fig:correlation}(c)-(d), we plot the Kapitza resistance against $\gamma$ and $\rho$, respectively. At small and large tilt angles, where the defect density is relatively small, there is a 
clear linear dependence of $R$ on both $\gamma$ and $\rho$. However, at intermediate tilt angles ($2\theta \approx 30\degree$), 
where the defect density is relatively large, the linear dependences become less clear, especially between $R$ and $\gamma$, 
which may indicate increased interactions between the defects. Overall, there is a stronger 
correlation between the Kapitza resistance and the defect density which is consistent with the idea of enhanced phonon scattering
with increasing $\rho$. 

\section{Summary and Conclusions}

In summary, we have employed an efficient multiscale modeling strategy based on the PFC approach and atomistic MD simulations to
systematically evaluate the Kapitza resistances in graphene grain boundaries for a wide range of tilt angles between adjacent grains. 
Strong correlations between the Kapitza resistance and the tilt angle, the grain boundary line tension, and the defect density are identified. 
Quantum effects, which have been ignored in previous studies, are found to be significant. By applying a mode-to-mode quantum correction method
based on spectral decomposition, we have demonstrated that good agreement between the classical molecular dynamics data 
and the quantum mechanical Landauer-B\"utticker method can be obtained. 

We emphasize that we have only considered suspended systems in this work. In a recent experimental work by 
Yasaei \textit{et al.}  \cite{yasaei2015}, Kapitza conductances (inverse of the Kapitza resistance) for a few supported 
(on SiN substrate) samples containing grain boundaries with different tilt angles were measured. The Kapitza conductances reported in 
this work are about one order of magnitude smaller than our quantum corrected values. This large discrepancy indicates that certain 
substrates may strongly affect heat transport across graphene grain boundaries and more work is needed to clarify this.

\section*{Acknowledgements}

This research has been supported by the Academy of Finland through its Centres of Excellence Program (Project No. 251748).
We acknowledge the computational resources provided by Aalto Science-IT project and Finland's IT Center for Science (CSC). 
K.A. acknowledges financial support from the Ministry of Science and Technology of Islamic Republic of Iran.
P.H. acknowledges financial support from the Foundation for Aalto University Science and Technology, and from the 
Vilho, Yrj\"o and Kalle V\"ais\"al\"a Foundation of the Finnish Academy of Science and Letters. 
Z.F. acknowledges the support of the National Natural Science Foundation of China (Grant No. 11404033). 
K.R.E. acknowledges financial support from the National Science Foundation under Grant No. DMR-1506634.

\bibliographystyle{elsarticle-num}

\bibliography{refs}

\end{document}